# MIMO decoding based on stochastic reconstruction from multiple projections

Amir Leshem[1,2]   and Jacob Goldberger[1]

### Abstract

Least squares (LS) fitting is one of the most fundamental techniques in science and engineering. It is used to estimate parameters from multiple noisy observations. In many problems the parameters are known a-priori to be bounded integer valued, or they come from a finite set of values on an arbitrary finite lattice. In this case finding the closest vector becomes NP-Hard problem. In this paper we propose a novel algorithm, the Tomographic Least Squares Decoder (TLSD), that not only solves the ILS problem, better than other sub-optimal techniques, but also is capable of providing the a-posteriori probability distribution for each element in the solution vector. The algorithm is based on reconstruction of the vector from multiple two-dimensional projections. The projections are carefully chosen to provide low computational complexity. Unlike other iterative techniques, such as the belief propagation, the proposed algorithm has ensured convergence. We also provide simulated experiments comparing the algorithm to other sub-optimal algorithms.

## I. INTRODUCTION

A multiple-input-multiple-output (MIMO) system is a communication system with $d$ transmit antennas and $p$ receive antennas. The tap gain from transmit antenna $i$ to receive antenna $j$ is denoted by $\mathbf{H}_{ij}$. In each use of the MIMO channel a signal vector $\mathbf{s} = (s_1, ..., s_d)^\top$ is independently selected from a set of constellation points $\mathcal{A}$ according to the data to be transmitted, so that $\mathbf{s} \in \mathcal{A}^d$. The received vector $\mathbf{x}$ is given by:

$$\mathbf{x} = \mathbf{Hs} + \mathbf{n} \tag{1}$$

[1]School of Engineering, Bar-Ilan University. [2]Dept. of EEMCS, Delft University of Technology

The vector $\mathbf{n}$ is an additive noise in which the noise components are assumed as zero mean, statistically independent Gaussians with a known variance $\sigma^2$. The channel matrix which is assumed to be known, comprises i.i.d. elements drawn from a (circularly symmetric zero-mean complex) normal distribution of unit variance. In the case where the MIMO linear system is complex-valued we use the standard method to translate it into an equivalent double-size real-valued representation that is obtained by considering the real and imaginary parts separately. The MIMO detection problem is then becomes finding the transmitted vector $\mathbf{s}$ given $\mathbf{H}$ and $\mathbf{x}$. The optimal maximum likelihood (ML) solution is:

$$\hat{\mathbf{s}} = \arg\min_{\mathbf{s} \in \mathcal{A}^d} \|\mathbf{H}\mathbf{s} - \mathbf{x}\|^2 \tag{2}$$

However, ML decoding has exponential computational complexity which makes it unfeasible when either the number of transmitted antennas or the constellation size are large. Actually, for a general $\mathbf{H}$, it is known to be NP-HARD both in the worst-case sense [1] as well as in the average sense [2]. It can be easily verified that the MIMO ML detection problem is equivalent to a least square lattice search problem that is known to be NP hard. A simple approximation is the zero-forcing (ZF) algorithm which is based on a linear decision ignoring the finite constellation constraint:

$$\hat{\mathbf{s}} = (\mathbf{H}^\top \mathbf{H})^{-1} \mathbf{H}^\top \mathbf{x} \tag{3}$$

and then, neglecting the correlation between the symbols, finding the closest constellation point for each symbol independently. This scheme performs poorly due to its inability to handle ill-conditioned channel matrix realizations. Somewhat better performance can be obtained by using a minimum mean square error (MMSE) filter instead of ZF on the un-constrained linear system:

$$\hat{\mathbf{s}} = (\mathbf{H}^\top \mathbf{H} + \sigma^2 I)^{-1} \mathbf{H}^\top \mathbf{x} \tag{4}$$

and then finding the closest lattice point in each component independently. A vast improvement over the linear approach can be achieved by using sequential decoding. This algorithm, known as MMSE V-BLAST or MMSE-SIC, has the best performance for this family of linear-based algorithms. However, there is a still a significant gap between the detection performance of the MMSE-SIC algorithm and the performance of the optimal ML detector. The complexity of all these algorithms is $O(p^3)$ where $p$ is the number of receive antennas (we assume $p \geq d$). These algorithms can also easily provide probabilistic (soft-decision) estimates for each symbol (or each bit).

Many alternative structures have been proposed to approach the ML detection performance. For example, the sphere decoding algorithm [3], approaches using the sequential Monte Carlo framework [4] and methods based on semidefinite relaxation [5], [6] have successfully been implemented. Although the detection schemes listed above have significantly reduced computational complexity, sphere decoding is still exponential in the average case [7]) and semidefinite relaxation is high-degree polynomial. Neither of these approaches can be easily used in real-world hardware architecture applications. Since these algorithms find the closest point in the lattice, it is not straight-forward to compute a-posteriori probabilities per symbol or per bit, which is required, when forward error correction is used (e.g., in communication applications), but it can be done with increased complexity of the sphere decoding procedure [8]. Thus, there is still a need for low complexity detection algorithms that can achieve good performance with low-order polynomial complexity, that are capable of providing per-bit likelihood ratios.

In this paper we propose a novel iterative technique, which we dub *Tomographic Least Squares Decoder* (TLSD), that is based on 2D projections followed by iterative optimization. The solution also allows us to provide a-posteriori probability distributions to each bit of each variable, something desirable in coded communication systems. Such probabilities are more complicated to evaluate using sphere decoding types of solution.

The paper proceeds as follows. In Section 2 we present the proposed TLSD algorithm. Experimental results are shown and discussed in Section 3.

## II. Tomographic Decoding of Constrained Linear Systems

In this section we present a novel polynomial time algorithm for solving the bounded integer least squares problem. The algorithm outperforms other reduced complexity algorithms with lower complexity. The algorithm has two important steps. The first step is translating the problem into a set of two-dimensional problems. The second step comprises of solving iteratively the two-dimensional problems by using data received from other two-dimensional problems. This is very similar to tomographic imaging, where an object is reconstructed from its projections on lower-dimensional subspaces. Hence we dub it *Tomographic Least Squares Decoder* (TLSD). The difference is that our object is discrete, and the data that is shared among the projections consists of probability distributions. The second step can also be interpreted as an instance of the incremental EM algorithm [9]. This will allow us to prove the convergence of the algorithm.

### A. The two-dimensional projections

Our approach can be viewed as a combination of a two-dimensional generalization of the ZF solution with optimal solution of the 2D subsystems obtained by this generalization.

Let $\mathbf{h}_1, ..., \mathbf{h}_d$ be the columns of $\mathbf{H}$ and for each $1 \leq i < j \leq d$ let $\mathbf{A}_{ij}$ be the matrix obtained from $\mathbf{H}$ by removing both $i$-th and $j$-th columns. It can be easily verified that the transformation:

$$\mathbf{P}_{ij} = \mathbf{I} - \mathbf{A}_{ij}(\mathbf{A}_{ij}^\top \mathbf{A}_{ij})^{-1} \mathbf{A}_{ij}^\top \tag{5}$$

is an orthogonal projection into the complement of the sub-space spanned by $\{\mathbf{h}_k | k \neq i, j\}$. Hence

$$\mathbf{P}_{ij}\mathbf{H}\mathbf{s} = \mathbf{P}_{ij} \sum_k \mathbf{h}_k s_k = \mathbf{P}_{ij}\mathbf{h}_i s_i + \mathbf{P}_{ij}\mathbf{h}_j s_j \tag{6}$$

Applying the linear transformation $\mathbf{P}_{ij}$ on both sides of the equation $\mathbf{H}\mathbf{s} + \mathbf{n} = \mathbf{x}$, yields a set of $p$ equations that depends only on the two variables $s_i$ and $s_j$:

$$\mathbf{P}_{ij}\mathbf{h}_i s_i + \mathbf{P}_{ij}\mathbf{h}_j s_j + \mathbf{P}_{ij}\mathbf{n} = \mathbf{P}_{ij}\mathbf{x} \tag{7}$$

Using the simplifying notation: $\mathbf{H}_{ij} = \mathbf{P}_{ij}[\mathbf{h}_i, \mathbf{h}_j]$, $\mathbf{n}_{ij} = \mathbf{P}_{ij}\mathbf{n}$ and $\mathbf{x}_{ij} = \mathbf{P}_{ij}\mathbf{x}$, Eq. (7) can be written as:

$$\mathbf{H}_{ij}[s_i, s_j]^\top + \mathbf{n}_{ij} = \mathbf{x}_{ij} \tag{8}$$

where $\mathbf{n}_{ij} \sim \mathcal{N}(0, \sigma^2 \mathbf{P}_{ij})$. The density function of $\mathbf{x}_{ij}$ is:

$$f_{ij}(\mathbf{x}_{ij}; s_i, s_j) = \frac{1}{2\pi\sigma^2} \exp(-\frac{1}{2\sigma^2} \|\mathbf{x}_{ij} - \mathbf{H}_{ij}[s_i, s_j]^\top\|^2) \tag{9}$$

Note that this is a two dimensional density function since the vector $\mathbf{x}_{ij}$ belongs to a two dimensional subspace spanned by $\mathbf{P}_{ij}\mathbf{h}_i$ and $\mathbf{P}_{ij}\mathbf{h}_j$. Furthermore, the orthogonal projection of an isotropic Gaussian variable is still isotropic in the projected space.

We have converted the original linear system into $\binom{d}{2}$ systems of sparse linear equations. If we take only non-overlapping projections (e.g. $\mathbf{P}_{12}, \mathbf{P}_{34}, ..., \mathbf{P}_{d-1,d}$) and solve the corresponding linear systems, it can be easily verified that we get exactly the linear ZF solution. Our approach is based on taking *all* the $\binom{d}{2}$ sparse systems. Due to the overlap between the projections, each of the solvers of the sub-problems provides information to the other solvers.

Ignoring the noise correlation between equation sets obtained by different projections, the likelihood function of $\mathbf{s} \in \mathcal{A}^d$, based on the sparse linear systems:

$$\mathbf{H}_{ij}[s_i, s_j]^\top + \mathbf{n}_{ij} = \mathbf{x}_{ij} \qquad , \qquad 1 \leq i < j \leq d \tag{10}$$

is:

$$f(\mathbf{x}; \mathbf{s}) = \prod_{i<j} f_{ij}(\mathbf{x}_{ij}; s_i, s_j) \tag{11}$$

$$= (\frac{1}{2\pi\sigma^2})^{\binom{d}{2}} \exp(-\frac{1}{2\sigma^2} \sum_{i<j} \|\mathbf{x}_{ij} - \mathbf{H}_{ij}[s_i, s_j]^\top\|^2)$$

Note that $f(\mathbf{x}; \mathbf{s})$ is not the precise likelihood function since we correlation between equations derived from different projections. Note however that all pairwise correlations are still captured by the relevant 2-D subproblems, so basically we only give up noise correlations of order 3 and higher.

Our goal now is finding the maximum-likelihood solution of the new system: $\hat{\mathbf{s}} = \arg\max_{\mathbf{s}} f(\mathbf{x}; \mathbf{s})$. In the next section we present an iterative method for maximizing $f(\mathbf{x}; \mathbf{s})$. The main point of this paper is that by applying the 2D projections we shift from the original likelihood function into a very similar function that is much easier to optimize. The sparsity of the new system makes $f(\mathbf{x}; \mathbf{s})$ a much smoother function than the original likelihood function. This smoothness enables applying an effective iterative search. A similar situation occurs in LDPC codes[10] where the sparsity of the parity-check matrix results in a smooth likelihood function.

### B. Iterative solution of the sparse problem

We have now converted the original linear system, into $\binom{d}{2}$ sets of sparse equations. The second step of our approach comprises of solving iteratively the two-dimensional problems by using data received from other two-dimensional problems.

Given an a-priori probability vector for $s_i, s_j$ we can now easily use $\mathbf{x}_{ij}$ to update these probabilities in a locally optimal way. Assume that for each $i = 1, ..., d$ we have an a-priori probability distribution on $s_i$ i.e., probability vectors $\boldsymbol{\theta}_i = (\theta_i(1), ..., \theta_i(M))$, where $\theta_i(k) = p(s_i = a_k)$, where $\mathcal{A} = \{a_1, ..., a_M\}$ is the finite symbol set. Given $\mathbf{x}_{ij}$ we can compute the a-posteriori probability for $s_i, s_j$ denoted by $\boldsymbol{\theta}_i^a, \boldsymbol{\theta}_j^a$ respectively and given by

$$\theta_i^a(k) \quad \propto \quad \theta_i(k) \sum_{\ell=1}^{M} \theta_j(\ell) D_{ij}(a_k, a_\ell) \tag{12}$$

$$\theta_j^a(\ell) \quad \propto \quad \theta_j(\ell) \sum_{k=1}^{M} \theta_i(k) D_{ij}(a_k, a_\ell)$$

where

$$D_{ij}(a_k, a_\ell) = f_{ij}(\mathbf{x}_{ij}; a_k, a_\ell) \tag{13}$$

$$= \frac{1}{2\pi\sigma^2} \exp(-\frac{1}{2\sigma^2} \|\mathbf{x}_{ij} - \mathbf{H}_{ij}[a_k, a_\ell]^\top\|^2)$$

and the notation $\propto$ indicates normalization of the vector to make it a distribution. We can now iterate the updates of $\boldsymbol{\theta}_i, i = 1, ..., d$, by choosing at each iteration a new pair $i < j$ and updating $\boldsymbol{\theta}_i, \boldsymbol{\theta}_j$. It can be shown that this is an instance of the EM algorithm.

To initialize the process we need a good choice of the a-priori probability vectors $\boldsymbol{\theta}_i$. This can be done for example using a soft version of the ZF solution. Let $\mathbf{P}_i$ be the ZF one-dimensional orthogonal projection into complement of the subspace spanned by $\{\mathbf{h}_k | k \neq i\}$. Then the initial parameter values are:

$$\theta_i(k) \propto \exp(-\frac{1}{2\sigma^2}\|\mathbf{P}_i\mathbf{h}_i a_k - \mathbf{P}_i\mathbf{x}\|^2) \tag{14}$$

### C. The TLSD algorithm

To decode an integer LS problem we perform the following: We first compute all the matrices $\mathbf{H}_{ij}$. This amounts to $\binom{d}{2}$ QR factorizations for each $i < j$. This has complexity $O\left(d^2p^3\right)$, but it is done once in the beginning of the decoding process. Now for each received vector $\mathbf{x}$ we first use a ZF receiver to generate the a-priori probability distributions $\theta_i, i = 1, \ldots, d$. This has complexity of $O(p^3)$, since the main problem is the computation of the ZF receiver. Computing the priors is $O(Md)$.

The next step is to compute for each two-dimensional vector of constellation points the metric using equation (13). This has complexity of $O\left(d^2M^2\right)$. This is done once for each received vector. Now we go over all vectors $\mathbf{x}_{ij}$ sequentially and update $\theta_i, \theta_j$ using (12). This is done until convergence is achieved, typically with few iterations. The overall complexity is $O(M^2N_{\text{iter}})$. After convergence the we obtain a-posteriori probabilities per symbol. A hard-decision solution is given by choosing for each $i = 1, ..., d$ the most probable symbol:

$$\hat{s}_i = \arg\max_{1 \leq k \leq M} \theta_i(k) \tag{15}$$

The algorithm-box in Table I summarizes the TLSD algorithm. The proposed TLSD algorithm is based only on two-dimensional subspaces. It is straightforward to improve the algorithm by using projections on higher dimensional subspace. This can have improved performance, and higher computational complexity. Finally we compare the likelihood of the solution vector with the likelihood of the MMSE-SIC solution and choose the one with higher likelihood. It turns out that since these algorithms use different type of information about the solution, that this improves the performance, especially for low SNR situations.

### III. SIMULATIONS

In this section we provide simulation results for the proposed detector over various uncoded MIMO systems. We assume a quasi-static fading channel with a frame length of 100. Under the assumption of block-fading channel model, the channel matrix $\mathbf{H}$ is constant for 100 channel uses. The channel matrix comprised i.i.d. elements drawn from a zero-mean normal distribution of unit variance. We have

---

Input: An integer LS problem: $\mathbf{Hs} + \mathbf{n} = \mathbf{x}$, a noise level $\sigma^2$

and a finite symbol set $\{a_1, ..., a_M\}$.

Initialization:

For $i = 1, ..., d$

initialize $\langle \theta_i(k) : a_k \in \mathbb{A} \rangle$ using zero-forcing (14).

For each pair $1 \le i < j \le d$

Compute the projection $\mathbf{P}_{ij}$ using Eq. (5) and compute:

$$D_{ij}(a_k, a_\ell) = \exp(-\tfrac{1}{2\sigma^2} \|\mathbf{P}_{ij}(\mathbf{x} - \mathbf{h}_i a_k - \mathbf{h}_j a_\ell)\|^2) \ .$$

End

---

Do until convergence

For $1 \le i < j \le d$

Update the distributions $\theta_i, \theta_j$:

$$\theta_i(k) \propto \theta_i(k) \sum_{\ell=1}^{M} \theta_j(\ell) D_{ij}(a_k, a_\ell)$$

$$\theta_j(\ell) \propto \theta_j(\ell) \sum_{k=1}^{M} \theta_i(k) D_{ij}(a_k, a_\ell)$$

End

End

For $i = 1, ..., d$

Choose $\hat{s}_i = \arg\max_k \theta_i(k)$.

---

used 10,000 realizations of channel matrix. This results in $10^6$ vector messages. The SNR is defined as $10 \log_{10}(E_b/N_0)$ where $E_b$ is the average received energy per symbol at each receiver antenna.

Fig. 1 shows the symbol error rate (SER) versus SNR for a $8 \times 8$ BPSK MIMO system. The performance of the TLSD method is compared to ML detection and to other linear suboptimal algorithms: the linear MMSE and the sequential MMSE V-BLAST. In all our experiments the number of TLSD iterations was limited to 10. It can be seen that the TLSD algorithm is significantly better than the MMSE-SIC at the same computational complexity. Fig. 2 depicts similar results for a $16 \times 16$ 4-PAM MIMO system. The TLSD decoder significantly outperforms the MMSE-SIC, while having comparable computational complexity.

## IV. CONCLUSIONS

Solving integer least squares problems is an important problem in many fields. We have proposed a novel technique based on tomographic principle of reconstruction from projections. We showed that the method always converges. Furthermore, the proposed method has good performance competitive to all other polynomial algorithms for solving the problem as demonstrated in simulations. Finally the method can be extended to provide a-posteriori probabilities per bit for use in coded communication systems or combined with sphere decoding, to improve its performance.

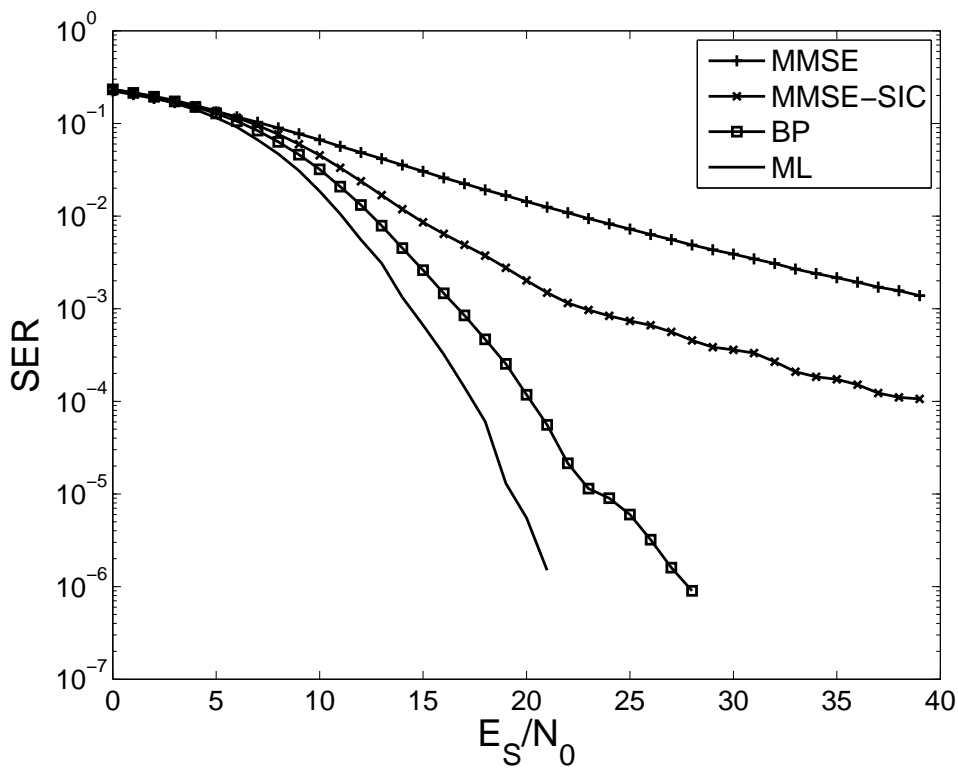

Fig. 1.   Results for $8 \times 8$ system, $\mathcal{A} = \{-1, 1\}$

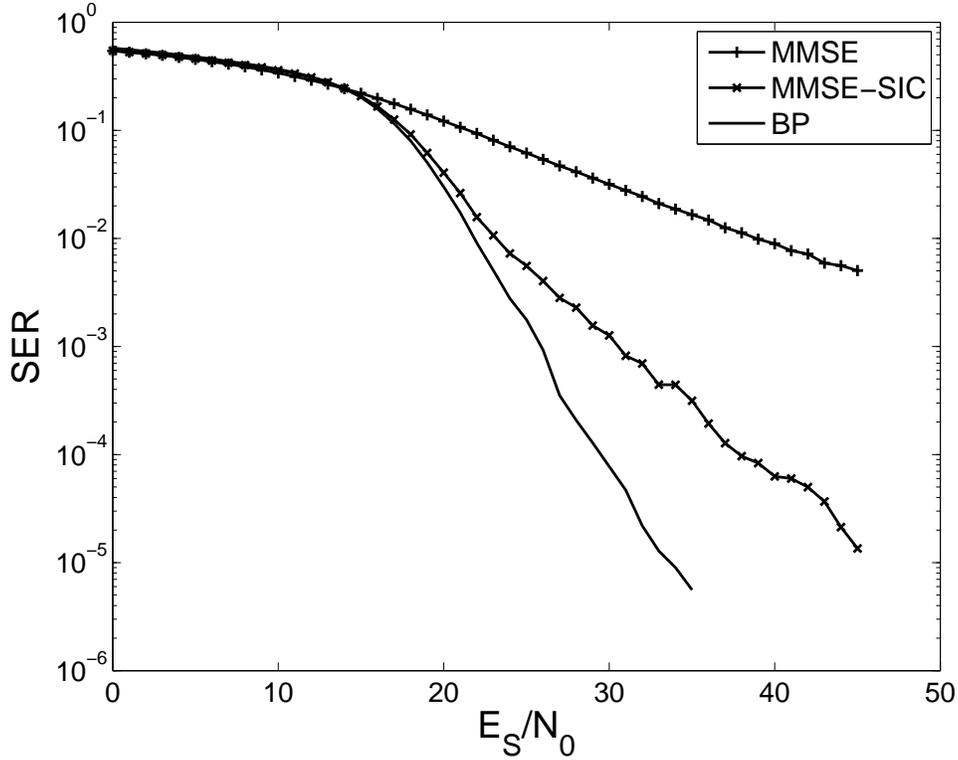

Fig. 2.   Results for $16 \times 16$ system, $\mathcal{A} = \{\pm 1, \pm 3\}$